\documentclass[lettersize,journal]{IEEEtran}

\normalsize

\usepackage{algorithmic}
\usepackage[caption=false,font=normalsize,labelfont=sf,textfont=sf]{subfig}
\usepackage{textcomp}
\usepackage{verbatim}
\usepackage{graphicx}
\def\BibTeX{{\rm B\kern-.05em{\sc i\kern-.025em b}\kern-.08em
    T\kern-.1667em\lower.7ex\hbox{E}\kern-.125emX}}
\usepackage{balance}

\usepackage{cite}

\usepackage[cmex10]{amsmath}
\usepackage{bbm}
\usepackage{array}

\usepackage{mdwmath}
\usepackage{mdwtab}
\usepackage{fixltx2e}
\usepackage{stfloats}
\usepackage{url}

\usepackage{comment}
\usepackage{times}
\usepackage{mathptmx}
\usepackage{newtxmath}
\usepackage[utf8]{inputenc}
\usepackage{dsfont}
\usepackage{graphicx}
\usepackage[english]{babel}

\usepackage{pstricks,bbm,bm,psfrag,pifont,times} 
\usepackage{hyperref}
\usepackage{cite}
\usepackage{color}
\usepackage[normalem]{ulem}

\DeclareMathOperator{\Tr}{Tr}
\def\Id{ {\mathbbm{1}} }
\def\ns{n_{\rm S} }
\def\WFH{{\rm WH}}
\def\HL{{\rm HL}}
\def\DWFH{{\rm DW}}
\def\p{{\rm p}}
\def\q{{\rm q}}

\makeatother

\usepackage{babel}

\begin{document}
\global\long\def\bra#1{\mathinner{\langle{#1}|}}%
\global\long\def\ket#1{\mathinner{|{#1}\rangle}}%
\global\long\def\braket#1{\mathinner{\langle{#1}\rangle}}%
\global\long\def\Id{{\mathbbm1}}%
\title{Employing weak-field homodyne detection for quantum communications}

\author{Michele~N.~Notarnicola and
        Stefano~Olivares 
\thanks{Manuscript received \today. 
{\em (Corresponding author: Stefano Olivares)}}
\thanks{Michele~N.~Notarnicola and Stefano~Olivares are with the Dipartimento di Fisica ``Aldo Pontremoli'', Università degli Studi di Milano, I-20133 Milano, Italy, and also with INFN, Sezione di Milano, I-20133 Milano, Italy (e-mail: stefano.olivares@fisica.unimi.it).}
}

\maketitle

\begin{abstract}
We investigate the role of weak-field homodyne (WF) measurement for quantum communications over a lossy bosonic channel with coherent state encoding. This kind of receiver employs photon-number resolving (PNR) detectors with finite resolution and low-intensity local oscillator. As a figure of merit, we consider the mutual information for a Gaussian input modulation. We prove an enhancement over Shannon capacity in the photon starved regime, obtained by exploiting information on the mean signal energy to suitably optimize the local oscillator intensity. Thereafter, we investigate the performance of non-Gaussian modulation, by considering a Gamma distribution of the energy of the encoded pulses, and achieve an increase in the information rate with respect to the Gaussian modulation case in the intermediate energy regime, being more accentuated for low values of the PNR resolution.
\end{abstract}

\begin{IEEEkeywords}
Communication channels, channel capacity.
\end{IEEEkeywords}

\IEEEpeerreviewmaketitle

\section{Introduction}

\IEEEPARstart{T}{he} transmission of classical information over a lossy channel is one of the fundamental problems in optical communication, widely addressed by both classical and quantum information theory \cite{Cover2006, Cariolaro2015}. 
In classical communications, a sender encodes a sequence of symbols into the quadratures of an optical field, while the receiver decodes it via coherent detection, implemented by single or double homodyne measurement \cite{Cariolaro2015, Holevo2011, Banaszek2020}. The maximum achievable information rate, referred to as the (classical) channel capacity, follows from the Shannon-Hartley theorem, and it is achieved by a Gaussian modulation at the input \cite{Holevo2011}. Nevertheless, in many different conditions, e.g. high-loss transmission, long-distance and deep-space communications, the probed signals are so attenuated that the particle-light nature of electromagnetic radiation emerges, thus a quantum description in terms of photons is strictly required \cite{Hemmati2011}. 

Approaching the problem in the quantum picture provides a wider and more fundamental perspective \cite{Gordon1962}. In fact, at the quantum limit, information is carried by quantum states of radiation, which need to be distinguished by the receiver in optimal way to maximize the achievable information rate. This allows the adoption of non-classical states of light, e.g. Fock states, as carriers \cite{Yamamoto1986, Caves1994} as well as implementation of unconventional detection schemes at the receiver's side, exhibiting higher sensitivity than conventional quadrature detection \cite{Chen2012, DiMario2019, Notarnicola2023:KB}. By considering general quantum encoding and measurements, quantum information theory leads to the ultimate information rate, the so-called Holevo capacity, outperforming the Shannon limit \cite{Holevo2011, Banaszek2020}. 
Remarkably, the Holevo limit is reached by Gaussian modulation of coherent states, describing radiation emitted by a stable laser, that is the same optimal encoding adopted in the classical case. On the contrary, the optimal detection scheme is a collective measurement, whose actual implementation is in general unknown. In turn, a challenging task is to find simpler individual measurements still guaranteeing a quantum advantage over classical capacities. In particular, direct detection (DD), i.e. photon-number measurement implemented by photon-number resolving (PNR) detectors, has been proved to beat coherent detection in the low energy regime, achieving a larger channel capacity \cite{Bowen1968, Shamai1990, Martinez2007, Cheraghchi2019, Lukanowski2021}.

Following this direction, a fascinating solution is represented by hybrid receivers, combining both the properties of discrete- and continuous- variable measurements. Two paradigmatic examples are provided by weak-field homodyne ($\WFH$) \cite{Donati2014, Thekkadath2020} and homodyne-like ($\HL$) detection \cite{Allevi2017, Olivares2019, Chesi2019}. Both of them are based on a scheme that mimics the quantum homodyne detection, but employing PNR detectors with finite resolution and low-intensity local oscillator. In this paper we refer to $\WFH$ detection if we measure the photon-number statistics of the two output beams, whereas if we probe the sole difference photocurrent we deal with $\HL$.

Furthermore, $\HL$ has already been investigated in the field of quantum communication, proving itself as a powerful resource for both coherent-state discrimination \cite{Notarnicola2023:HYNORE, Notarnicola2023:FF, Notarnicola2024:PHN} and continuous-variable quantum key distribution \cite{Cattaneo2018}. In this paper, we address the performance of $\WFH$, $\HL$ and double weak-field homodyne ($\DWFH$) detection for the transmission of classical information over a lossy bosonic channel. We consider modulation of coherent states at the source and compute the resulting information rate under two configurations: at first, we perform Gaussian modulation, as for the Shannon and Holevo limits; then we lift the Gaussian hypothesis and propose a Gamma distribution for the energy of the coherent pulses as an example of non-Gaussian sampling. We prove Gaussian modulation to be preferable in the photon starved limit, in which regime weak-field measurements beat Shannon capacity. Otherwise, for intermediate values of the received energy, we show that a suitable non-Gaussian distribution enhances the information rate, whilst in the high energy limit Gaussian sampling returns to be optimal.

The structure of the paper is the following. At first, in Sec.~\ref{sec: OptComm} we briefly review the main features of quantum optical communications, then in Sec.~\ref{sec: WFH} we present the schemes of the proposed weak-field measurements. Thereafter, in Sec.~\ref{sec: Mutualnfo} we address the role of $\WFH$, $\HL$ and $\DWFH$ for quantum communications by considering Gaussian modulation at the source, while in Sec.~\ref{sec: NonGauss} we investigate the enhancement brought by non-Gaussian distributions. Finally, in Sec.~\ref{sec: Concl} we draw some conclusions.

\section{Fundamentals of optical communications}\label{sec: OptComm}
In an optical communication protocol, a sender, Alice, handles a classical source, described by a random variable $X$ whose possible values $x$ are generated with probability $p_A(x)$. The input symbols $x$ are then encoded into optical signals located in temporal slots of duration $B^{-1}$, $B$ being the slot rate characterizing the width of the spectrum in the frequency domain. At the quantum limit, each pulse is described by (possibly mixed) a quantum state $\rho_x$ with mean energy, i.e. mean number of photons, equal to $\bar{n}= P/(B \hbar \omega)$, $P$, $\hbar$ and $\omega$ being the input power, Planck's reduced constant and the carrier angular frequency, respectively \cite{Banaszek2020}. 
Throughout the paper we consider a narrowband scenario, i.e. $2\pi B \ll \omega$, where the encoding process is performed onto laser pulses.
Thereafter, the signal is injected into a quantum channel, modelled as a completely positive trace preserving map ${\cal E}$, until reaching the receiver, Bob, who will infer the value of symbol $x$. This procedure is realized by performing a suitable quantum measurement, described by a positive-operator valued measurement (POVM) $\{\Pi_y\}_y$, $\Pi_y \ge 0$, $\sum_y \Pi_y = \Id$, retrieving an outcome $y$ associated with a random variable $Y$ correlated to $X$ \cite{Holevo2011, Cariolaro2015, Banaszek2020}.
In a classical description in the absence of memory effects, the whole protocol is fully characterized by the conditional probability distribution $p_{B|A}(y|x)$, and the overall distribution of $Y$ reads $p_B(y)=\sum_x p_A(x) p_{B|A}(y|x)$ \cite{Cover2006}. On the contrary, in the quantum regime the conditional distribution follows from the Born rule \cite{Paris2012}, namely $p_{B|A}(y|x)= \Tr[{\cal E}(\rho_x) \Pi_y]$, thus for each choice of the POVM at Bob's side we define a different classical channel, mapping the input variable $X$ into its counterpart $Y$. 

For a particular quantum measurement, the amount of information about $X$ extractable from $Y$ is equal to the mutual information:
\begin{eqnarray}
I(X:Y) = H(Y) - H(Y|X) \, ,
\end{eqnarray}
expressed in bits per time slot, i.e. per channel use. In the former expression, $H(Y)= {\sf H}[p_B(y)]$ and $H(Y|X)= \sum_x p_A(x) \times {\sf H}[p_{B|A}(y|x)]$ are the average and conditional Shannon entropies, respectively, where ${\sf H}[q(z)]=-\sum_z q(z) \log_2 q(z)$ is the Shannon entropy of distribution $q(z)$ \cite{Holevo2011}. Mutual information optimized over all possible input distributions yields the channel capacity:
\begin{eqnarray}\label{eq:Capacity}
C = \max_{p_A(x)} I(X:Y) \, ,
\end{eqnarray}
representing the maximum amount of extractable information given the input ensemble $\{\rho_x\}_x$ and the POVM $\{\Pi_y\}_y$. We remind that the supremum in Eq.~(\ref{eq:Capacity}) should be computed under a suitable constraint on the average input power, i.e. on the mean photon number per symbol. 

Beyond classical limits, the ultimate channel capacity is obtained by further optimizing the mutual information over all quantum measurements and state ensembles. By resorting to the Holevo theorem \cite{Holevo1973}, this ultimate transmission rate is provided by the so-called Holevo capacity, equal to \cite{Holevo1998, Giovannetti2014}:
\begin{eqnarray}\label{Holevo:3}
C_{\rm H} = \max_{\{\rho_x, p_A(x)\}} \left\{ {\sf S} [{\cal E}(\bar{\rho})] - \sum_x p_A(x) {\sf S} [{\cal E}(\rho_x)] \right\}\, ,
\end{eqnarray}
where $\bar{\rho}= \sum_x p_A(x) \rho_x$ is the average state received by Bob and ${\sf S}[\rho]=-\Tr[\rho \log_2 \rho]$ is the von Neumann entropy of state $\rho$ \cite{Holevo2011}.

In optical communication at the quantum limit, the single-mode
radiation field is described by a bosonic operator $a$, satisfying the canonical commutation relations, $[a, a^\dagger ]=1$. The canonical field quadratures are then obtained as $q=a+a^\dagger$ and $p= i (a^\dagger -a)$, expressed in shot-noise units. We note that the two operators do not commute, as $[q,p]=2i$, therefore it is not possible to simultaneously measure them with maximum precision, according to Heisenberg's uncertainty relation.
Moreover, in this scenario, typical information carriers are provided by coherent states, describing radiation emitted by laser sources. A coherent state $|\alpha\rangle$, characterized by a complex amplitude $\alpha \in \mathbb{C}$, is a superposition of $n$-photon states, namely $|\alpha\rangle= e^{-|\alpha^2|/2} \sum_n \alpha^n/\sqrt{n!}|n\rangle$, having mean photon number or, equivalently, energy $|\alpha|^2$ \cite{Olivares2021}.
A milestone example of a quantum channel is the lossy bosonic channel, modeling transmission inside optical fibers in the absence of additive Gaussian excess noise \cite{Giovannetti2004, Holevo2001}. The channel is characterized by a transmissivity $\tau \le 1$, quantifying signal attenuation. In particular, after pure-loss transmission a coherent state remains coherent, albeit with a reduced amplitude, $|\alpha\rangle \rightarrow |\sqrt{\tau}\alpha\rangle$. 
Thus, Bob probes a rescaled coherent state ensemble and the resulting channel capacities only depend on the mean received energy per time slot,
\begin{eqnarray}
\ns= \tau \bar{n} \, ,
\end{eqnarray}
$\bar{n}$ being the average input energy at Alice's side. In turn, we compute the information rate as for a lossless channel with reduced input modulation energy, performing the substitution $\bar{n} \rightarrow \ns$. 

\begin{figure}[tb!]
\centerline{\includegraphics[width=0.9\columnwidth]{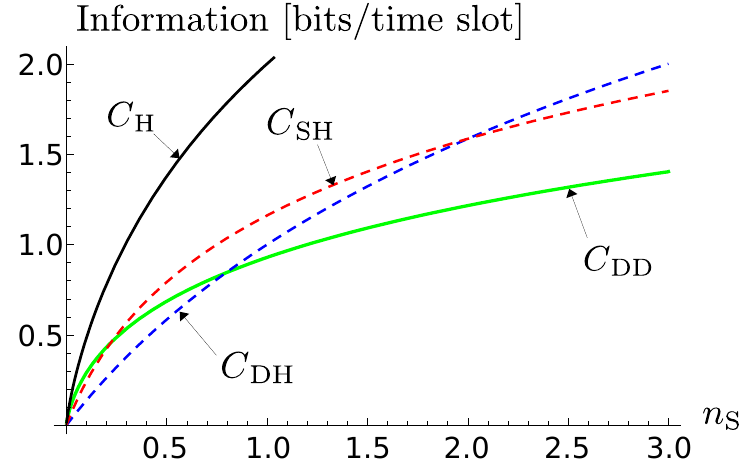} }
\caption{Plot of the Shannon capacities $C_{\rm r}$, $\rm r=SH,DH$, in Eq.~(\ref{eq:ShannonCapacity}), the Holevo capacity $C_{\rm H}$ in~(\ref{eq:HolevoCapacity}), and the DD upper bound $C_{\rm DD}$ in~(\ref{eq:DDcapacity}) as a function of the mean received energy $\ns$. In the low energy regime $\ns <1$, the DTP capacity $C_{\rm DD}$ outperforms the Shannon limit.}\label{fig:01-Fundamentals}
\end{figure}

The well known Shannon capacities arising from the the Shannon-Hartley theorem \cite{Shannon1949} are obtained via measurement of either one or both canonical field quadratures $q$ and $p$, achieved by single- (SH) or double-homodyne (DH) detection, respectively \cite{Olivares2021, Banaszek2020}.  The corresponding capacity is reached with Gaussian modulation of the coherent state amplitude $\alpha=x_A+i y_A$. For the homodyne case, we set $y_A \equiv 0$ and the uni-variate modulation $p_A(x_A)={\cal N}_{\sigma^2}(x_A)$, where
\begin{eqnarray}\label{eq:Gauss}
{\cal N}_{\sigma^2}(x) = \frac{\exp\big[-x^2/(2\sigma^2)\big]}{\sqrt{2\pi \sigma^2}} \, ,
\end{eqnarray}
whereas in the presence of double-homodyne detection we adopt a bi-variate modulation, namely $p_A(x_A,y_A)= {\cal N}_{\sigma^2}(x_A){\cal N}_{\sigma^2}(y_A)$. In turn, the average quantum state at Bob's side contains $\ns= \sigma^2$  and $\ns= 2\sigma^2$ mean photons, respectively. Then, the Shannon capacities read:
\begin{IEEEeqnarray}{rCl}\label{eq:ShannonCapacity}
\IEEEyesnumber \IEEEyessubnumber*
C_{\rm SH} &=&\frac12 \log_2 \big(1+4\ns \big)\, , \label{eq:ShannonH} \\[1ex] 
C_{\rm DH} &=& \log_2 \big(1+\ns \big)\, , \label{eq:ShannonDH}
\end{IEEEeqnarray}
and are reported in Fig.~\ref{fig:01-Fundamentals} as functions of the mean received signal energy.
We note that $C_{\rm SH} \ge C_{\rm DH}$ for $\ns\le 2$, whereas for higher energy the double-homodyne capacity becomes larger than the single quadrature one. In fact, the joint measurement of both the non-commuting $q$ and $p$ operators, corresponding to the double-homodyne detection, introduces an ineludible excess noise equal to the shot-noise vacuum fluctuations, thus reducing the available signal-to-noise ratio (SNR) \cite{Banaszek2020}. In turn, there is a tradeoff between this reduced SNR and the increase of accessible information due to the bi-variate signal modulation, such that if the signal energy is sufficiently low single homodyne detection becomes preferable.

Remarkably, also the Holevo capacity (\ref{Holevo:3}) is reached by Gaussian modulation of coherent states \cite{Giovannetti2004, Banaszek2020}, and reads:
\begin{eqnarray}\label{eq:HolevoCapacity}
C_{\rm H} = g(\ns)\, ,
\end{eqnarray}
where $g(x)=(x+1)\log_2(x+1) -x \log_2 x$.
However, the optimal POVM achieving~(\ref{eq:HolevoCapacity}) is a collective measurement, operating simultaneously on multiple time slots, whose associated detection scheme is still unknown \cite{Holevo2011, Banaszek2020, Cariolaro2015}. 
Thus, the interest has been directed to either design simple collective measurements approximating the Holevo capacity in particular energy regimes \cite{Guha2011, Banaszek2017}, or to find feasible suboptimal individual measurement performed on single time slots \cite{Bowen1968, Shamai1990}.

Within the latter scenario, an intriguing solution to overcome the Shannon limit is provided by DD. In literature, the quantum channel employing coherent state encoding and DD at the output is referred to as discrete-time Poisson (DTP) channel \cite{Shamai1990, Martinez2007, Cheraghchi2019, Lukanowski2021}. Differently from the previous cases, here Gaussian modulation is no longer the optimal choice, and to evaluate the DTP capacity a numerical calculation is required \cite{Lukanowski2021}. However, simpler bounds on the actual capacity have been established. In particular, in 2007 Martinez obtained a lower bound by considering a Gamma distribution for the energy of the carrier pulses, together with an upper bound based on the duality property of constrained convex optimization \cite{Martinez2007}. More recently, a tighter upper bound was established by Cheraghchi et al., equal to: \cite{Cheraghchi2019}
\begin{IEEEeqnarray}{rCl}\label{eq:DDcapacity}
C_{\rm DD} &=& \ns \log_2 \Bigg[ \frac{1 + (1 + e^{1 +\gamma})\ns  + 2 \ns^2}{e^{1 +\gamma}\ns  +2 \ns^2} \Bigg] \, \nonumber \\[1ex]
&+&\log_2 \Bigg[ 1 + \frac{1}{\sqrt{2e}} \Bigg(\sqrt{\frac{1 + (1 + e^{1 +\gamma})\ns  + 2 \ns^2}{1+\ns}} -1 \Bigg) \Bigg]\, ,
\end{IEEEeqnarray}
where $\gamma \approx 0.5772$ is the Euler-Mascheroni constant. 
We can see in Fig.~\ref{fig:01-Fundamentals} that DD is suboptimal with respect to quadrature detection in the high energy regime, as $C_{\rm DD} \le C_{\rm r}$, for $\ns \ge~n_{\rm r}$, $\rm r=SH,DH$, where $n_{\rm SH} \approx 0.22$ and $n_{\rm DH} \approx 0.79$. On the contrary, for low $\ns$, we have $C_{\rm DD} > C_{\rm r}$ and, remarkably, the exact DTP capacity numerically evaluated in \cite{Lukanowski2021} beats the Shannon limit too, reducing the gap with the Holevo capacity in the photon starved regime $\ns \ll 1$. 

The previous analysis suggest that possible improvements may be obtained by a hybrid receiver combining both the advantages of photon-number resolving detection, implementing the DD scheme, and quadrature measurements. To this aim, weak-field homodyne detection, presented in the next section, represents an intriguing solution.

\section{Weak-field homodyne detection}\label{sec: WFH}
\begin{figure}
\centerline{\includegraphics[width=0.95\columnwidth]{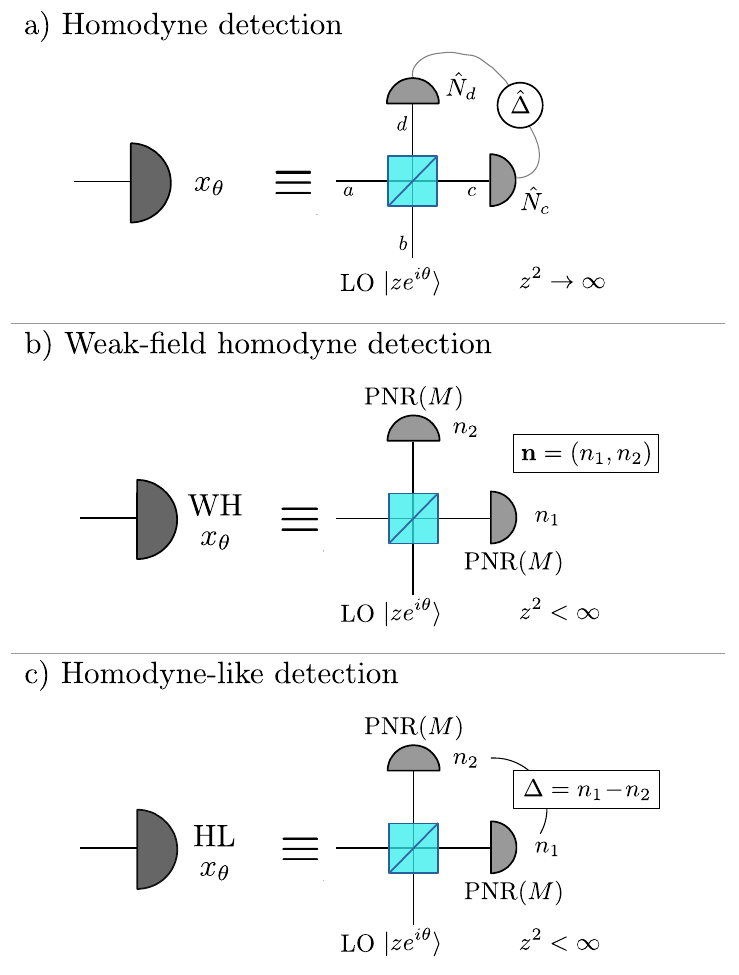} }
\caption{(a) Setup of homodyne detection of quadrature $x_\theta = a e^{-i\theta} + a^\dag e^{i\theta}$. The signal interferes with a high-intensity LO $|z e^{i\theta}\rangle$ at a balanced beam splitter, with $z^2 \rightarrow \infty$; thereafter, photon-number detection is performed on both branches and the difference photocurrent is considered. (b) Scheme of weak-field homodyne ($\WFH$) detection employing a low LO, $z^2 < \infty$, and PNR($M$) detection. The outcome is a pair of integer values ${\bf n}=(n_1,n_2)$, $n_{1(2)}=0,\ldots, M$. (c) Homodyne-like ($\HL$) detection, namely a $\WFH$ scheme where we only consider the difference photocurrent $\Delta=-M,\ldots, M$ at the output.
}\label{fig:02-WFH}
\end{figure}

Conventional coherent receivers in optical communications are based on homodyne detection, namely a phase-sensitive scheme to perform measurement of the field quadrature $x_{\theta}= \cos \theta \, q +\sin\theta\, p$, $0\le \theta < \pi$. Its implementation can be summarized as in Fig.~\ref{fig:02-WFH} (a). At first, the incoming signal is mixed at a balanced beam splitter with a local oscillator (LO) excited in the coherent state $|z e^{i \theta}\rangle$, $z>0$; thereafter photon-number detection is performed on both the output beams. We note that this requires both the signal and LO fields to be frequency-matched \cite{Banaszek2020}.

At the quantum limit, we describe the input optical modes associated with signal and LO by two bosonic operator $a$ and $b$, respectively, while
performing photon-number detection on the output modes $c$ and $d$ corresponds to measure the two Hermitian operators $\hat{N}_c= c^\dagger c$ and $\hat{N}_d=d^\dagger d$.
Ultimately, we evaluate the difference photocurrent $\hat{\Delta} = \hat{N}_{c}-\hat{N}_{d}$.
In the Heisenberg picture, namely by applying the modes transformations at a balanced beam splitter $a \to c = (a+b)/\sqrt{2}$ and $b \to d = (b-a)/\sqrt{2}$, we have \cite{Olivares2021}:
\begin{eqnarray}
\hat{\Delta} =  a b^\dagger + a^\dagger b \,  .\label{eq:DiffPhoto}
\end{eqnarray}
Therefore, considering the state $|z e^{i \theta}\rangle$ of the LO as fixed, we have:
 \begin{align}\label{eq:DiffP}
\langle z e^{i \theta}| \hat{\Delta} |z e^{i \theta}\rangle = z  \left(a e^{-i \theta} + a^\dagger e^{i\theta} \right) = z \, x_\theta \, .
\end{align}
In turn, measuring the rescaled photocurrent $\hat{\Delta}/z$ may provide an indirect measurement of the quadratures of the input field. If we compute, however,  the expectation value of $\hat{\Delta}^n$ we have:
\begin{eqnarray}\label{delta:moment}
\frac{\langle z e^{i \theta}| \hat\Delta^n | z e^{i \theta}\rangle}{z^n}
= x_\theta^n + \gamma_{\theta}^{(n)}(a,a^\dag;z) \, ,
\end{eqnarray}
where $\gamma_{\theta}^{(n)}(a,a^\dag;z)$ is a function of both the field operators $a$ and $a^\dag$ and the LO intensity, to be explicitly calculated; for instance, $\gamma_{\theta}^{(1)}(a,a^\dag;z)=0$, $\gamma_{\theta}^{(2)}(a,a^\dag;z)=a^\dag a/z^2$, $\gamma_{\theta}^{(3)}(a,a^\dag;z)=(3 a^\dag x_\theta\, a + x_\theta)/z^2$ and
\begin{eqnarray}\label{delta:moment:4}
\gamma_{\theta}^{(4)}(a,a^\dag)=\frac{3 (a^\dag)^2 a^2+ a^\dag a}{z^4} + \frac{6 a^\dag x_\theta^2\, a + 4(x_\theta^2-1)+1}{z^2} \, .
\end{eqnarray}
Therefore, it is clear that the actual measurement of the quadrature is achieved only in the limit $z^2 \to \infty$ in which the measurement of the moments $(\hat\Delta/z)^n$ corresponds to measure $x_\theta^n$ \cite{Olivares2020}.

Moreover, in practical realizations, in the presence of high LO, photon-number measurement is implemented by p-i-n photodiodes, namely photodetectors generating macroscopic photocurrents proportional to the intensity of the incoming light.

Now, an intriguing solution is to investigate the performance of the homodyne scheme beyond the high-LO limit, when $z^2 < \infty$, in which case the former setup does not implement quadrature detection anymore.~

We can identify two possible scenarios. The first one is referred to as weak-field homodyne ($\WFH$) detection, whose scheme is reported in Fig.~\ref{fig:02-WFH}(b): the conventional p-i-n photodiodes of the standard homodyne detection scheme are replaced by PNR detectors \cite{Donati2014, Thekkadath2020}. The use of PNR detectors gives access to the two local photon-number statistics and allows
jointly probing both the wake- and particle-like properties of the field.
As a matter of fact, realistic PNR detectors have a finite resolution $M < \infty$, i.e. they can only resolve up to $M$ photons and, consequently, a low-intensity LO is required. To remark this point, throughout the paper we refer to them as PNR$(M)$ detectors. 
We describe PNR$(M)$ detection by a POVM with $M+1$ possible outcomes, $\{\Pi_0, \Pi_1, \ldots, \Pi_M \}$, with:
\begin{align}
\Pi_n = 
\left\{\begin{array}{ll} 
 | n \rangle\langle n| &\mbox{if}~ n=0,\ldots, M-1 \, , \\[2ex]
 \displaystyle \Id - \sum_{j=0}^{M-1} | j \rangle\langle j| &\mbox{if}~ n=M .
\end{array}
\right.
\end{align}
In particular, PNR($1$) detectors corresponds to on-off detectors, whereas DD, implemented by ideal photodetectors, is requires PNR($M$) detectors with $M=\infty$ \cite{Notarnicola2023:HYNORE}.

Accordingly, $\WFH$ detection returns a pair of integer outcomes ${\bf n}=(n_1,n_2)$, $n_{1(2)}=0,\ldots, M$. Given an input coherent state $|\alpha\rangle$, $\alpha \in \mathbb{C}$, the resulting probability distribution reads:
\begin{align}\label{eq:WFHdistr}
P^{(\theta)}_{\WFH}({\bf n}|\alpha)=  q_{n_1}\big(\mu_{+}(\alpha;\theta)\big) q_{n_2}\big(\mu_{-}(\alpha;\theta)\big) \,,
\end{align}
where 
\begin{align}\label{eq:mupm}
\mu_{\pm}(\alpha;\theta)= \frac{\left|\alpha \pm z e^{i\theta}\right|^2}{2}\, ,
\end{align}
is the mean energy on the two output branches, respectively, and
\begin{align}\label{eq:PnPNRM}
    q_{n}(\mu) =
    \left\{\begin{array}{l l}
    {\displaystyle e^{-\mu} \ \frac{\mu^{n}}{n!}}  & \mbox{if}~n<M \ , \\[2ex]
    {\displaystyle 1- e^{-\mu} \sum_{j=0}^{M-1} \frac{\mu^{j}}{j!}} & \mbox{if}~n= M \ ,
    \end{array}
    \right.
\end{align}
is the probability of obtaining the outcome $n$ from PNR$(M)$ detection, namely a truncated Poisson distribution.

In the second scenario we consider the $\WFH$ scheme where we evaluate the difference $\Delta$ between the number of photons measured at output beams. Due to the clear analogy with the standard homodyne detection, we refer to this configuration as homodyne-like ($\HL$) detection, depicted in Fig.~\ref{fig:02-WFH}(c) \cite{Allevi2017, Olivares2019, Chesi2019}. The probability of obtaining the value $\Delta=n_1-n_2$, with $-M \le \Delta \le M$, becomes:
\begin{align}\label{eq:probDelta}
S^{(\theta)}(\Delta|\alpha)= \sum_{n_1,n_2=0}^{M} q_{n_1}\big(\mu_{+}(\alpha;\theta)\big) q_{n_2}\big(\mu_{-}(\alpha;\theta)\big) \, \delta_{n_1-n_2,\Delta} \,,
\end{align}
$\delta_{j,k}$ being the Kronecker delta. In particular, when $M=\infty$, $\Delta$ represents the difference between two Poisson variables, thus Eq.~(\ref{eq:probDelta}) approaches a Skellam distribution \cite{Notarnicola2023:HYNORE}.

\begin{figure}
\centerline{\includegraphics[width=0.9\columnwidth]{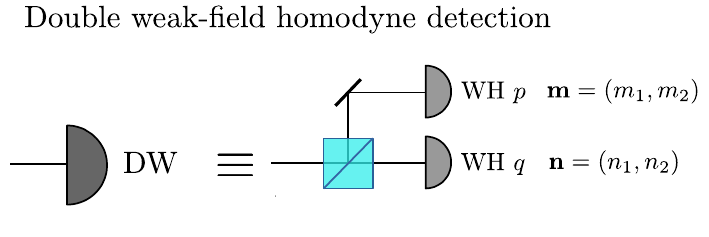} }
\caption{Setup of double weak-field homodyne ($\DWFH$) detection. The incoming signal is divided into two parts at a balanced beam splitter, on which joint $\WFH$ measurement of both quadrature is performed. The measurement outcome is provided by the tuple of integers ${\bf K}=({\bf n},{\bf m})=(n_1,n_2;m_1,m_2)$, $n_{1(2)}, m_{1(2)}=0,\ldots, M$.}\label{fig:03-DWFH}
\end{figure}

We can straightforwardly extend the previous analysis to the joint quadrature measurements and
introduce a double weak-field homodyne ($\DWFH$) measurement via the scheme in Fig.~\ref{fig:03-DWFH}. Now we split the incoming signal in two parts thanks to a balanced beam splitter and, then, implement joint $\WFH$ detection of quadrature $q$ (corresponding to $\theta=0$) and $p$ ($\theta=\pi/2$) on the transmitted and reflected branch, and we choose the same LO intensity $z^2$ for both setups. The two performed measurements lead to outcomes ${\bf n}=(n_1,n_2)$ and ${\bf m}= (m_1,m_2)$, respectively. Ultimately, the outcome of $\DWFH$ consists in the tuple ${\bf K}=({\bf n},{\bf m})=(n_1,n_2,m_1,m_2)$. The resulting $\DWFH$ probability distribution for a coherent input $|\alpha\rangle$ reads:
\begin{align}\label{eq:DWFHdistr}
P_{\DWFH} ({\bf K}|\alpha)=P^{(0)}_{\WFH}\Bigg({\bf n}\, \Big|\frac{\alpha}{\sqrt{2}}\Bigg) P^{(\pi/2)}_{\WFH}\Bigg({\bf m}\, \Big|\frac{\alpha}{\sqrt{2}}\Bigg) \, .
\end{align}

\section{Application to quantum communications}\label{sec: Mutualnfo}

Given the previous considerations, we now investigate the role of $\WFH$, $\HL$ and $\DWFH$ measurements for quantum communication over a lossy channel of transmissivity $\tau$ and a coherent state ensemble at the input. 
In the present paper, we compute the mutual information for different examples of input distributions, with the intent of showing, as a proof-of-principle, whether or not weak-field measurements overcome Shannon capacities and guarantee a quantum advantage.

Here we consider a Gaussian modulation of the input coherent state amplitudes, whereas in the next section 
we will investigate the non-Gaussian modulation case, by proposing a suitable input distribution inspired on the existing results for DTP channels \cite{Martinez2007}. The calculation of the channel capacity is beyond the scope of this work: it is an open problem for more advanced studies, since it needs a rigorous numerical treatment exploiting the Blahut-Arimoto algorithm in its continuous-variable version \cite{Lukanowski2021, Blahut1972, Arimoto1972}.

In our analysis we first address the performance of single quadrature detection schemes, and compare the mutual information associated with both $\WFH$ and $\HL$ measurements. Thereafter, we tackle the case of double weak-field quadrature measurement, and investigate the enhancement in the information rate brought by $\DWFH$ detection.

\subsection{Single quadrature detection: $\WFH$ vs $\HL$}\label{sec: SingleComp}
\begin{figure}
\centerline{\includegraphics[width=0.9\columnwidth]{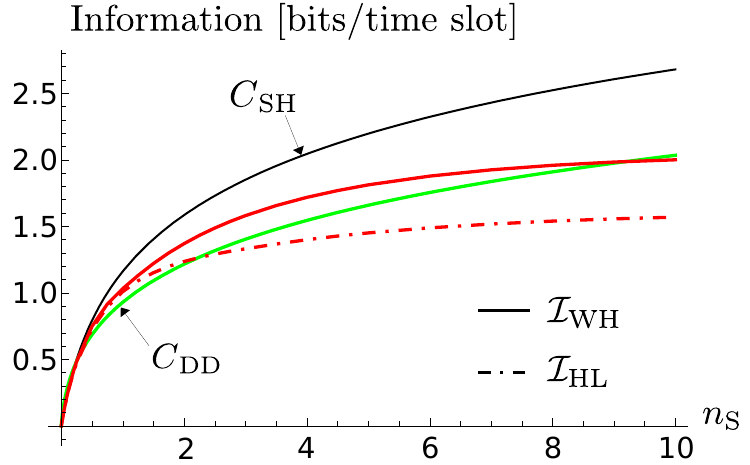} }\vspace*{2ex}
\centerline{\includegraphics[width=0.9\columnwidth]{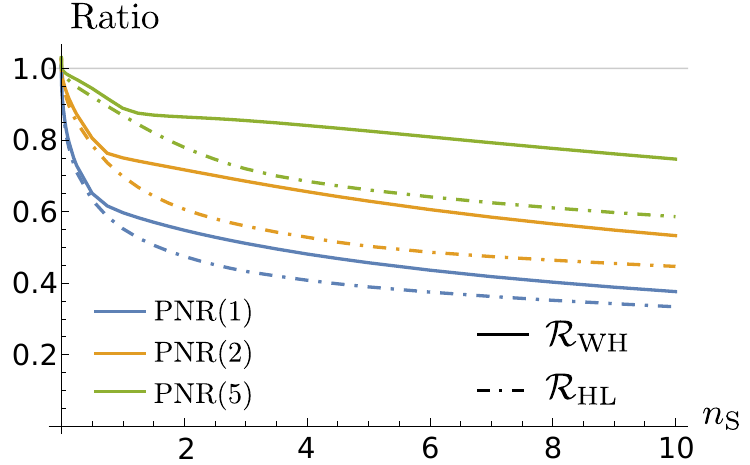} }
\caption{(Top) Plot of the mutual information ${\cal I}_{\p}$, $\p=\WFH,\HL$, achieved with Gaussian modulation, as a function of the mean received energy $\ns$. The PNR resolution is $M=5$. $C_{\rm SH}$ and $C_{\rm DD}$ refers to the Shannon homodyne capacity~(\ref{eq:ShannonH}) and the DD upper bound~(\ref{eq:DDcapacity}), respectively. (Bottom) Plot of the ratio ${\cal R}_{\p} = {\cal I}_{\p} / C_{\rm SH}$, $\p=\WFH,\HL$, as a function of the mean received energy $\ns$ for different PNR resolution $M$.}
\label{fig:04-MInfoS}
\end{figure}

In the single quadrature detection Alice encodes the information only on the real amplitude $x_A\in \mathbb{R}$ of the coherent states. We can consider a Gaussian modulation of the sole $q$ quadrature, namely $p_A(x_A)={\cal N}_{\sigma^2}(x_A)$, see Eq.~(\ref{eq:Gauss}), while $y_A$ is set to $0$, 
such that the probed received energy is equal to $\ns=\sigma^2$.
At the receiver's side we compare the performance of two alternative cases in which Bob performs either $\WFH$ or $\HL$ measurement of the same quadrature $q$.

In the $\WFH$ case, Bob, after detection, retrieves a pair of integer outcomes ${\bf n}_B=(n_1,n_2)$, $n_{1(2)}=0,\ldots, M$, $M$ being the photon-number resolution. The corresponding conditional and unconditional probabilities read $p_{B|A}({\bf n}_B|x_A)= P^{(0)}_{\WFH}({\bf n}_B|\alpha=x_A)$ in Eq.~(\ref{eq:WFHdistr}) and
\begin{eqnarray}
p_B({\bf n}_B)= \int_{\mathbb{R}} dx_A\,  {\cal N}_{\sigma^2}(x_A) \, p_{B|A}({\bf n}_B|x_A) \, ,
\end{eqnarray}
respectively, and the mutual information is obtained as:
\begin{eqnarray}\label{eq:InfoWFH}
{\cal I}_{\WFH} = \max_{z>0} \, {\cal I}_{\WFH}(z) \, ,
\end{eqnarray}
where
\begin{IEEEeqnarray}{lCl}
{\cal I}_{\WFH} (z)&=& {\sf H} \left[p_B({\bf n}_B)\right] \nonumber \\[1ex]
&&- \int_{\mathbb{R}} dx_A \, {\cal N}_{\sigma^2}(x_A) \, {\sf H} \left[p_{B|A}({\bf n}_B|x_A)\right]  \, ,
\end{IEEEeqnarray}
and maximization is performed over the LO amplitude of the $\WFH$ scheme.

Instead, in the presence of $\HL$ measurement, Bob measures only the difference photocurrent $-M\le \Delta_B \le M$ without knowing the local statistics of the two PNR($M$) detectors, thus the conditional conditional and unconditional probabilities become $p_{B|A}(\Delta_B|x_A)= S^{(0)}(\Delta_B|\alpha=x_A)$ in~(\ref{eq:probDelta}) and
\begin{eqnarray}
p_B(\Delta_B)= \int_{\mathbb{R}} dx_A\,  {\cal N}_{\sigma^2}(x_A) \, p_{B|A}(\Delta_B|x_A) \, ,
\end{eqnarray}
respectively, leading to:
\begin{eqnarray}\label{eq:InfoHL}
{\cal I}_{\HL} = \max_{z>0} \, {\cal I}_{\HL}(z) \, ,
\end{eqnarray}
with
\begin{IEEEeqnarray}{lCl}
{\cal I}_{\HL} (z)&=& {\sf H} \left[p_B(\Delta_B)\right] \nonumber \\[1ex]
&&- \int_{\mathbb{R}} dx_A \, {\cal N}_{\sigma^2}(x_A) \, {\sf H} \left[p_{B|A}(\Delta_B|x_A)\right]  \, .
\end{IEEEeqnarray}

Plots of ${\cal I}_\p$, $\p=\WFH, \HL$, are reported in Fig.~\ref{fig:04-MInfoS} (top panel) as a function of $\ns$ for fixed PNR resolution $M=5$, compared to the Shannon capacity $C_{\rm SH}$ and the DD upper bound $C_{\rm DD}$ in Eq.s~(\ref{eq:ShannonH}) and ~(\ref{eq:DDcapacity}), respectively. 
The behavior is analogous for all values of the resolution $M$. We see that, for $\ns \gtrsim 10^{-3}$, weak-field measurements perform worse than homodyne detection, and the difference between ${\cal I}_\p$ and $C_{\rm SH}$ increases for larger $\ns$, as the finite resolution becomes progressively insufficient to resolve all the photons reaching the detector.
However, we have ${\cal I}_\p \ge C_{\rm DD}$ in the limit of $\ns > n_{\rm SH} \approx 0.22$ and sufficiently large $M$, therefore, even in the presence of a finite resolution, $\WFH$ and $\HL$ outperform the DD capacity, 
achieved by PNR$(\infty)$ detection.
Moreover, in all energy regimes, $\HL$ leads to a lower information rate than $\WFH$. This is a consequence of the data processing inequality \cite{Cover2006}. In fact, the conditional $\HL$ probability~(\ref{eq:probDelta}) is obtained by combining the $\WFH$ probabilities~(\ref{eq:WFHdistr}) via a suitable post-processing strategy, being a local operation not able to increase the accessible information on the encoded signal.

To quantify the information reduction with respect to the Shannon limit, we compute the ratio
\begin{eqnarray}\label{eq:RatioSH}
{\cal R}_{\p} = \frac{{\cal I}_{\p}}{C_{\rm SH}}  \, , \qquad \p=\WFH, \HL \, ,
\end{eqnarray}
showed in the bottom panel of Fig.~\ref{fig:04-MInfoS}. Consistently with the previous considerations, this ratio, for both $\WFH$ and $\HL$, is a decreasing function of the received energy $\ns$, such that ${\cal R}_{\HL} \le {\cal R}_{\WFH}$. In particular, $\WFH$ enhances the mutual information up to the $25 \%$ in the low energy regime, whilst in the high energy limit the two ratios approach each other, namely ${\cal R}_{\HL} \approx {\cal R}_{\WFH}$ for $\ns \gg 1$.

We conclude that, in the context of quantum communications, $\WFH$ provides a preferable choice to $\HL$ and measuring the local PNR statistics is necessary to enhance the achievable information rate. 

\subsection{Photon starved regime}

Remarkably, a different performance of weak-field measurements is observed in the photon starved regime $\ns\ll 1$, where the mean received energy is much less than $1$, being a typical scenario in deep-space communications \cite{Hemmati2011}. In this regime, both $\WFH$ and $\HL$ provide a quantum advantage over the Shannon homodyne capacity.

To assess the performance of the schemes under investigation, it is convenient to express the mutual information as ${\cal I}_\p= \ns \, {\sf P}_\p$, where
\begin{eqnarray}\label{eq:PIE}
{\sf P}_\p = \frac{{\cal I}_\p}{\ns} \, , \qquad \p= \WFH, \HL \, ,
\end{eqnarray}
is the photon information efficiency (PIE), representing the number of bits per received photon \cite{Banaszek2020, Dolinar2011}. 
The PIE provides a measurement of the efficiency of information transfer per unit energy, being of particular relevance when the received energy is so low that Bob receives at most one photon per each time slot.

\begin{figure}
\centerline{\includegraphics[width=0.9\columnwidth]{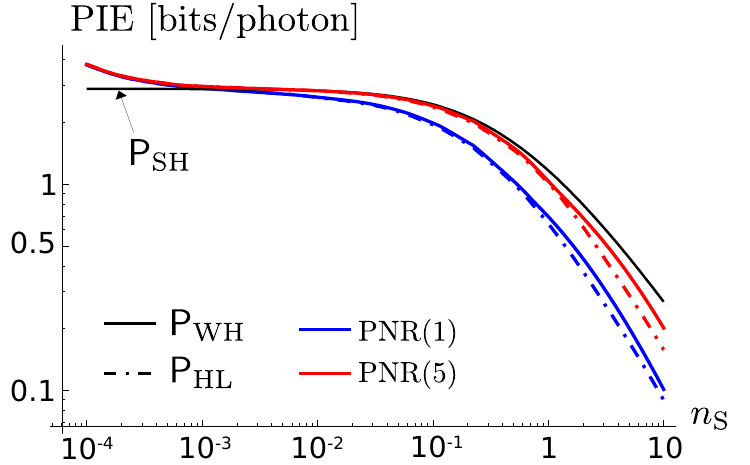} }\vspace*{2ex}
\centerline{\includegraphics[width=0.9\columnwidth]{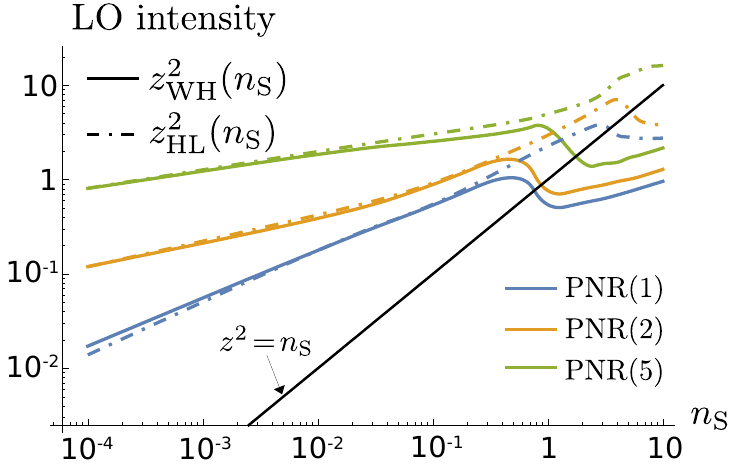} }
\caption{(Top) Log-log plot of the PIE ${\sf P}_{\p}$, $\p=\WFH, \HL$, as a function of the mean received energy $\ns$ for different PNR resolution $M$. Weak-field measurements outperforms the Shannon capacities in the photon starved regime $\ns \lesssim 10^{-3}$. 
(Bottom) Log-log plot of the optimized LO $z^2_{\p}(\ns)$, $\p=\WFH, \HL$ as a function of $\ns$ for different PNR resolution.
}\label{fig:05-PIESingle}
\end{figure}

Plots of ${\sf P}_\p$ are reported in Fig.~\ref{fig:05-PIESingle} (top panel) for different $M$, compared to the Shannon PIE, ${\sf P}_{\rm SH}= C_{\rm SH}/\ns$. As we can see, for $\ns \lesssim 10^{-3}$,
both $\WFH$ and $\HL$ outperform the homodyne capacity even in the presence of on/off detection, namely, PNR($1$), while increasing the resolution $M$ enlarges the region where we observe an enhancement.
These results can be explained considering the role of the LO intensity.
Differently from the homodyne limit $z^2 \rightarrow \infty$, in the presence of weak-field measurements, the signal is mixed with a finite LO, whose optimized intensity $z^2_{\p}(\ns)$, $\p= \WFH, \HL$, is reported in the bottom panel of Fig.~\ref{fig:05-PIESingle}, being a function of the energy $\ns$.

In the photon starved limit, we have $\ns \ll z^2_{\p}(\ns) \lesssim 1$, and for the Gaussian coherent state ensemble under investigation, the average statistical moments of the weak-field distributions are equal to
$\langle \hat{\Delta}^n/z^n \rangle = \langle q^n \rangle + \gamma^{(n)}(\ns;z)$, where the correction terms read $\gamma^{(1)}(\ns;z)=\gamma^{(3)}(\ns;z)=0$ and, as one can obtain from Eq.~(\ref{delta:moment}):
\begin{IEEEeqnarray}{lCl}
\IEEEyesnumber \IEEEyessubnumber*
\gamma^{(2)}(\ns;z) = \frac{\ns}{z^2} \, ,\\[1ex]
\gamma^{(4)}(\ns;z) = \frac{\ns(3\ns+1)}{z^4}+ \frac{3 \ns^2 + 8\ns}{z^2}
+ \frac{1}{z^2} \, .
\end{IEEEeqnarray}
Remarkably, we note that $\gamma^{(4)}(\ns;z)$ contains a constant term equal to $1/z^2$ independent of the signal and vanishing only in the high-LO limit.
As a consequence, in the weak-field scenario, we can suitably choose the LO intensity as a function of the mean signal energy, i.e. $z^2=z^2(\ns)$, to enhance the homodyne setup, optimize the statistical moments and, ultimately, obtain a probability distribution containing more information than the sole quadrature detection.
In particular, the results in Fig.~\ref{fig:05-PIESingle} (bottom panel) suggest that the optimal LO should satisfy the following two conditions. The former is $z^2(\ns) \gg \ns$, such that $\gamma^{(2)}(\ns;z) \ll 1$ and the variance of the distribution is not larger than conventional homodyne; the latter is to adopt a mesoscopic LO, $z^2(\ns) \lesssim 1$ to have $\gamma^{(4)}(\ns;z) = 1/z^2(\ns)+ O(\ns/z^2(\ns)) \ne 0$, making the shape of both the $\WFH$ and $\HL$ statistics significantly different than the homodyne (Gaussian) probability.

In the photon starved regime, the numerically optimized LO $z^2_{\p}(\ns)$ satisfies both conditions, whist for $\ns \gtrsim 10^{-3}$, they are not fulfilled anymore and the advantage over the Shannon capacity is lost.
Furthermore, the prevalence of $\WFH$ over $\HL$ due to the data processing inequality is maintained, as ${\sf P}_{\WFH} \gtrsim {\sf P}_{\HL}$ for all $\ns$.


\subsection{Double weak-field quadrature detection}\label{sec: DoubleComp}

\begin{figure}
\centerline{\includegraphics[width=0.9\columnwidth]{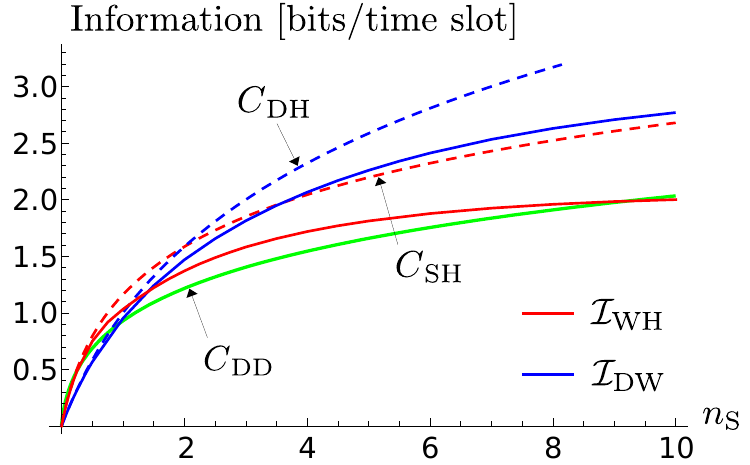} }\vspace*{2ex}
\centerline{\includegraphics[width=0.9\columnwidth]{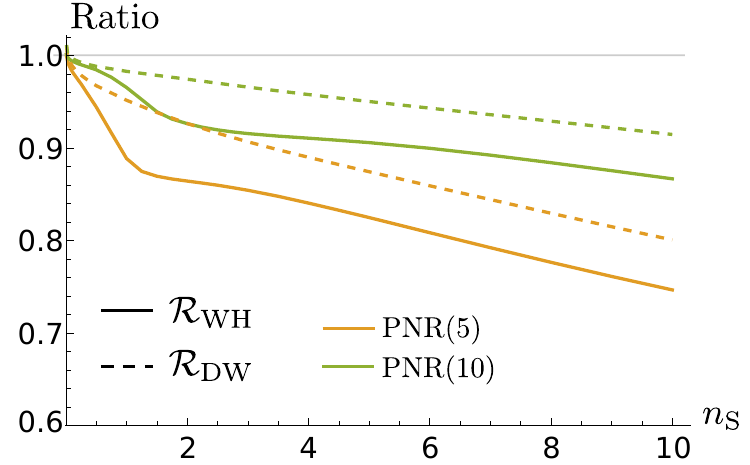} }
\caption{(Top) Plot of the mutual information ${\cal I}_{\q}$, $\q=\WFH,\DWFH$, achieved with Gaussian modulation, as a function of the mean received energy $\ns$. The PNR resolution is $M=5$. $C_{\rm SH}$, $C_{\rm DH}$ and $C_{\rm DD}$ refers to the Shannon homodyne and double-homodyne capacities~(\ref{eq:ShannonH}) and~(\ref{eq:ShannonDH}), and the DD upper bound~(\ref{eq:DDcapacity}), respectively. (Bottom) Plot of the ratio ${\cal R}_{\q}$, $\q=\WFH, \DWFH$, as a function of the mean received energy $\ns$ for different PNR resolution $M$.}\label{fig:06-MInfo}
\end{figure}

In the former subsection, we proved $\WFH$ to outperform $\HL$, showing that collecting the local PNR statistics of the output beams of a homodyne setup is a more powerful method than probing the sole difference photocurrent. Given this considerations, in the following we pick up $\WFH$ detection as the most appropriate single quadrature weak-field scheme. In turn, we now extend the analysis to double quadrature measurement, considering $\DWFH$ detection and computing the corresponding information rate.

In the presence of $\DWFH$, Alice implements a bi-variate Gaussian modulation $p_A(x_A,y_A)={\cal N}_{\sigma^2}(x_A){\cal N}_{\sigma^2}(y_A)$, and Bob retrieves the four-valued tuple ${\bf K}_B=(n_1,n_2,m_1,m_2)$, $n_{1(2)}, m_{1(2)}=0,\ldots, M$. The average received energy is $\ns=2\sigma^2$, as both quadrature amplitudes are modulated, and the probability distributions probed by Bob are $p_{B|A}({\bf K}_B|x_A,y_A)= P_{\DWFH}({\bf K}_B|\alpha=x_A+i y_A)$, reported in~(\ref{eq:DWFHdistr}), and
\begin{IEEEeqnarray}{lCl}
p_B({\bf K}_B)&=& \int_{\mathbb{R}^2} dx_A dy_A \, {\cal N}_{\sigma^2}(x_A){\cal N}_{\sigma^2}(y_A) \, \times \nonumber \\
&& \hspace{1.5cm} p_{B|A}({\bf K}_B|x_A,y_A) \, .
\end{IEEEeqnarray}
Ultimately, the mutual information is equal to:
\begin{eqnarray}\label{eq:InfoDWFH}
{\cal I}_{\DWFH} = \max_{z>0} \, {\cal I}_{\DWFH} (z) \, ,
\end{eqnarray}
where
\begin{IEEEeqnarray}{lCl}
{\cal I}_{\DWFH} (z)&=& {\sf H} \left[p_B({\bf K}_B)\right] \nonumber \\[1ex]
&&- \int_{\mathbb{R}^2} dx_Ady_A \, {\cal N}_{\sigma^2}(x_A){\cal N}_{\sigma^2}(y_A) \, \times \nonumber \\[1ex]
&&\hspace{1.5cm} {\sf H} \left[p_{B|A}({\bf K}_B|x_A,y_A)\right]  \, .
\end{IEEEeqnarray}

Plots of both ${\cal I}_{\DWFH}$ and ${\cal I}_{\WFH}$ are reported in Fig.~\ref{fig:06-MInfo} (top panel) as a function of $\ns$ for fixed PNR resolution $M=5$, compared to the Shannon and DD capacities~(\ref{eq:ShannonCapacity}) and ~(\ref{eq:DDcapacity}), respectively. 
As for conventional quadrature measurements, we have ${\cal I}_{\DWFH} \le {\cal I}_{\WFH}$ for $\ns \le n_{\rm W} <2$, $n_{\rm W}$ being an increasing function of the resolution $M$, approaching $2$ in the limit $M\gg 1$. This can be viewed as a consequence of the excess noise introduced by the joint quadrature detection. Moreover, like the single quadrature case, in the non-starved energy regime, $\DWFH$ leads to a reduced mutual information with respect to the Shannon limit $C_{\rm DH}$. We then consider the ratio

\begin{eqnarray}
{\cal R}_{\DWFH} = \frac{{\cal I}_{\DWFH}}{C_{\rm DH}} \, ,
\end{eqnarray}
depicted in Fig.~\ref{fig:06-MInfo} (bottom panel) together with the $\WFH$-detection ratio ${\cal R}_{\WFH}$ in Eq.~(\ref{eq:RatioSH}). For a given $M$, we see that ${\cal R}_{\WFH} \le {\cal R}_{\DWFH}$, thus the information reduction in the presence of a double quadrature weak-field measurement is lower than the single quadrature case. As expected, ${\cal R}_{\q}$, $\q=\WFH, \DWFH$, increase for higher resolution, and, for $\ns\le 10$, a PNR($10$) detector is sufficient to maintain both the two ratios above $\approx 90\%$.

\begin{figure}
\centerline{\includegraphics[width=0.9\columnwidth]{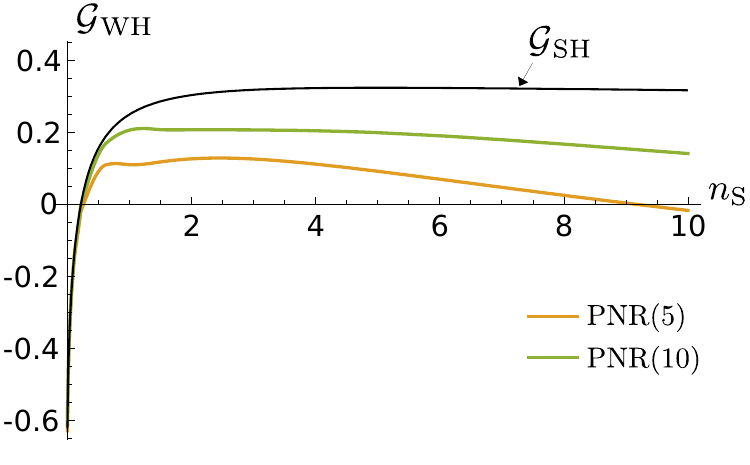} }\vspace*{2ex}
\centerline{\includegraphics[width=0.9\columnwidth]{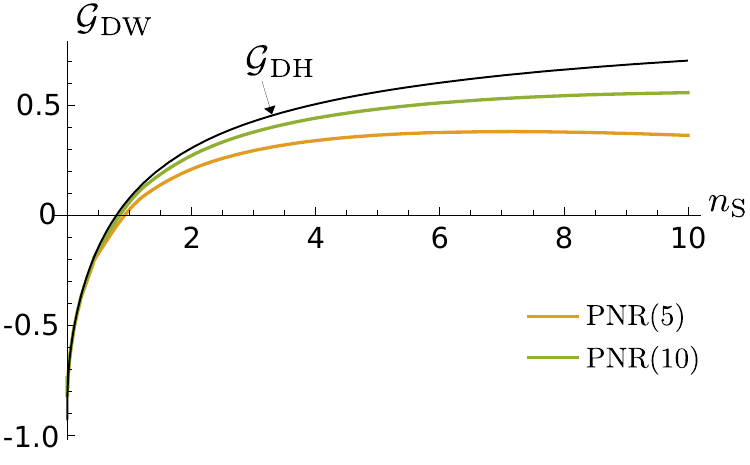} }
\caption{Plot of the gain ${\cal G}_{\WFH}$ (top panel) and ${\cal G}_{\DWFH}$ (bottom panel) as a function of the mean received energy $\ns$ for different PNR resolution $M$. ${\cal G}_{\rm SH}$ and ${\cal G}_{\rm DH}$ refers to the gain over DD achieved by single and double homodyne detection, respectively.}\label{fig:07-Gain}
\end{figure}

Moreover, despite their sub-optimality, in the limit $\ns > n_{\rm r}$, $\rm r=SH,DH$, $\WFH$ and $\DWFH$ with sufficiently high resolution outperform the DTP capacity, respectively.
To this aim, we introduce the gain over the DD upper bound, namely
\begin{eqnarray}
{\cal G}_\q= \frac{{\cal I}_{\q}}{C_{\rm DD}} -1 \, , \qquad \q=\WFH, \DWFH \, ,
\end{eqnarray}
plotted in the top and bottom panels of Fig.~\ref{fig:07-Gain} together with ${\cal G}_{\rm r}= C_{\rm r}/C_{\rm DD}-1$, $\rm r=SH,DH$, corresponding to the gain achieved by single and double homodyne detection, respectively. Consistently with the previous considerations, ${\cal G}_\q$ is not a monotonous function of $\ns$: it reaches a maximum value at a finite energy and then decreases due to the finite PNR resolution $M$. Furthermore, increasing the PNR resolution closes the gap with conventional quadrature detection. 
For instance, fixing a resolution $M=5$ guarantees a gain up to the $12\%$ for $\WFH$ measurement and $37\%$ for the $\DWFH$ case, whereas for $M=10$ the maximum achievable gain further increases to $20\%$ and $56\%$, respectively. 

\begin{figure}
\centerline{\includegraphics[width=0.9\columnwidth]{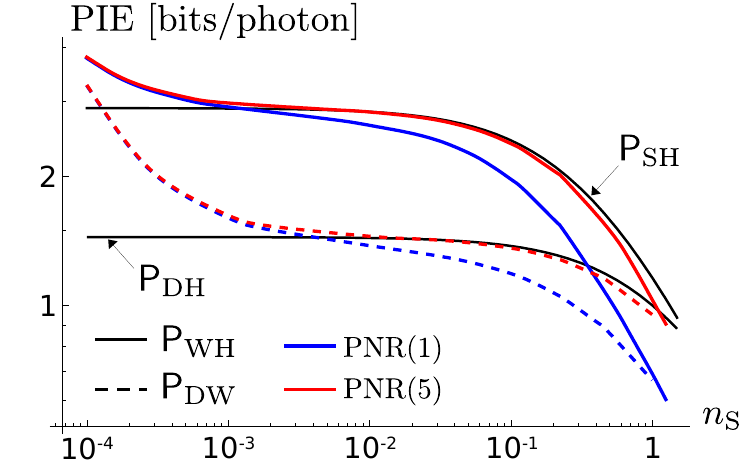} }
\caption{Log-log plot of the PIE ${\sf P}_{\q}$, $\q=\WFH,\DWFH$, as a function of the mean received energy $\ns$ for different PNR resolution $M$.}\label{fig:08-PIE}
\end{figure}

Finally, we investigate the accessible information rate in photon starved regime. We compute the double weak-field PIE
\begin{eqnarray}
{\sf P}_{\DWFH} = \frac{{\cal I}_{\DWFH}}{\ns} \, ,
\end{eqnarray}
and compare it to the $\WFH$-detection PIE in~(\ref{eq:PIE}).
Plots of ${\sf P}_\q$, $\q=\WFH,\DWFH$, are reported in Fig.~\ref{fig:08-PIE} for different $M$. Like the single quadrature scenario, for $\ns \lesssim 10^{-3}$, $\DWFH$ beats the double-homodyne capacity, i.e. ${\sf P}_{\DWFH} \ge {\sf P}_{\rm DH}$. 
In this regime $\WFH$ is preferable than $\DWFH$, as ${\sf P}_{\WFH} \ge {\sf P}_{\DWFH}$, but, remarkably, for sufficiently low $\ns$, ${\sf P}_{\DWFH}> {\sf P}_{\rm SH}$, and $\DWFH$ beats single homodyne detection too.
On the contrary, beyond the photon starved limit, for $\ns \gtrsim 10^{-3}$ both ${\sf P}_\q$, $ \q= \WFH, \DWFH$, decrease below the corresponding Shannon limits, retrieving the scenario discussed before.

\section{Beyond Gaussian modulation \label{sec: NonGauss}}

\begin{figure}
\centerline{\includegraphics[width=0.9\columnwidth]{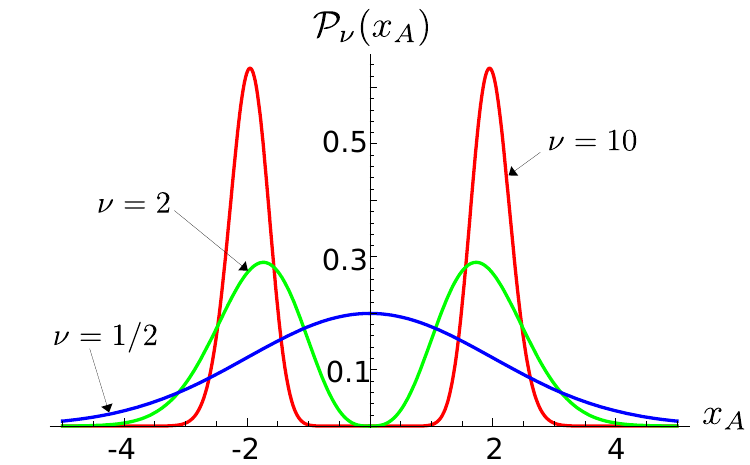} }
\caption{Plot of the non-Gaussian distribution ${\cal P}_\nu (x_A)$ as a function of $x_A$ for $\ns=4$ and different values of the free parameter $\nu$. When $\nu=1/2$ we retrieve the Gaussian probability ${\cal N}_{\ns} (x_A)$, while, for $\nu\rightarrow\infty$, ${\cal P}_\nu(x_A)$ approaches a double Dirac delta distribution.}\label{fig:09-GammaProb}
\end{figure}

\begin{figure}
\centerline{\includegraphics[width=0.9\columnwidth]{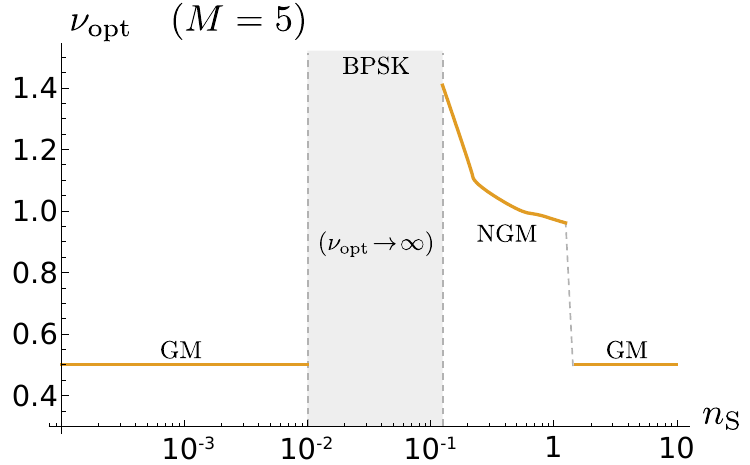} }
\caption{Log-linear plot of the optimized modulation parameter $\nu_{\rm opt}$ of the Gamma distribution as a function of the mean received energy $\ns$ for PNR$(5)$ detectors. Gaussian modulation (GM) is retrieved for $\nu=1/2$. On the contrary, $1/2<\nu_{\rm opt}<\infty$ realizes non-Gaussian modulation (NGM) and the limit $\nu_{\rm opt}\rightarrow \infty$ corresponds to the BPSK case. }\label{fig:10-nuopt}
\end{figure}

In the analysis conducted so far, we computed the information rate for weak-field measurements by assuming a Gaussian distribution at the input, proved as the optimal choice for the Shannon and Holevo capacities. However, $\WFH$ and $\DWFH$ are non-Gaussian measurements, thus reasonably Gaussian modulation (GM) will not be optimal and there should exists a suitable non-Gaussian distribution of the coherent state amplitudes leading to a higher mutual information. To this aim, we took inspiration from the research on DTP capacity and propose a paradigmatic example of non-Gaussian modulation (NGM) suggested by Martinez in \cite{Martinez2007}.
Dealing with the problem of bounding the DTP capacity, Martinez considered a Gamma modulation for the energy of the encoded pulses, providing a tight lower bound to the actual capacity of a DD scheme with coherent state ensembles \cite{Cheraghchi2019, Lukanowski2021}. 

We adopt a similar approach in the presence of weak-field measurements and, for the sake of simplicity, we only consider the single quadrature scenario with $\WFH$ detection. 
In this case, Alice generates coherent states with real amplitude $x_A\in\mathbb{R}$, whose corresponding energy $\epsilon=x_A^2$ is drawn from a Gamma distribution, characterized by the free parameter $\nu\ge 0$, namely:
\begin{eqnarray}\label{eq:Gamma}
p_\nu(\epsilon)= \frac{\nu^{\nu}}{\Gamma(\nu) \, \ns^{\nu}} \, \epsilon^{\nu-1} \exp\left(-\nu \frac{\epsilon}{\ns}\right) \, ,
\end{eqnarray}
where $\ns$ is the mean received energy and $\Gamma(\nu)$ is the Euler Gamma function. Once sampled the energy value $\epsilon$, she prepares a coherent pulse with amplitude $x_A= \pm \sqrt{\epsilon}$, whose sign is chosen at random with equal probability. The corresponding distribution for the coherent amplitude $\widetilde{p}_A(x_A)= {\cal P}_\nu (x_A)$ is then obtained as \cite{Hogg2019}:
\begin{IEEEeqnarray}{lCl}\label{eq:GammaxA}
{\cal P}_\nu (x_A)&=& \frac12 p_\nu(\epsilon) \frac{d\epsilon}{dx_A}\Bigg|_{\epsilon=x_A^2}  \nonumber \\[1ex]
&=& \frac{\nu^{\nu}}{\Gamma(\nu) \, \ns^{\nu}} \, \left(x_A^2\right)^{\frac{2\nu-1}{2}} \exp\left(-\nu \frac{x_A^2}{\ns}\right) \, ,
\end{IEEEeqnarray}
being non-singular for $\nu \ge 1/2$. 
As shown in Fig.~\ref{fig:09-GammaProb}, we identify three scenarios:
\begin{enumerate}
\item $\nu=1/2$ : in this case we retrieve the GM case, as ${\cal P}_\nu (x_A)={\cal N}_{\ns}(x_A)$; \vspace{.25cm}

\item $1/2<\nu <\infty$: now Eq.~(\ref{eq:GammaxA}) becomes a symmetric double peaked distribution centered in $x_{\rm \pm}= \pm \sqrt{(2\nu-1)\ns/(2\nu)}$ and such that  ${\cal P}_\nu (0)=0$; \vspace{.25cm}

\item $\nu \rightarrow \infty$: the amplitude probability becomes singular, approaching the double Dirac delta distribution ${\cal P}_\nu (x_A) \approx \left[\delta(x-\sqrt{\ns}) + \delta(x+\sqrt{\ns})\right]/2$: in fact, for large $\nu$, Eq.~(\ref{eq:Gamma}) converges to a normal distribution, which in the limit $\nu \rightarrow \infty$ approaches the Dirac delta $\delta(\epsilon-\ns)$ \cite{Hogg2019}.
\end{enumerate}
We remark that case $3)$ corresponds to a binary phase-shift keying (BPSK) modulation where Alice generates one of the two coherent states $|\pm \sqrt{\ns}\rangle$ with equal probability \cite{Notarnicola2023:HYNORE, Notarnicola2023:FF, Cariolaro2015}.
We also note that Martinez lower bound to the DTP capacity was obtained with the choice $\nu=1/2$, corresponding to the Gaussian coherent state ensemble \cite{Martinez2007}.

\begin{figure}
\centerline{\includegraphics[width=0.9\columnwidth]{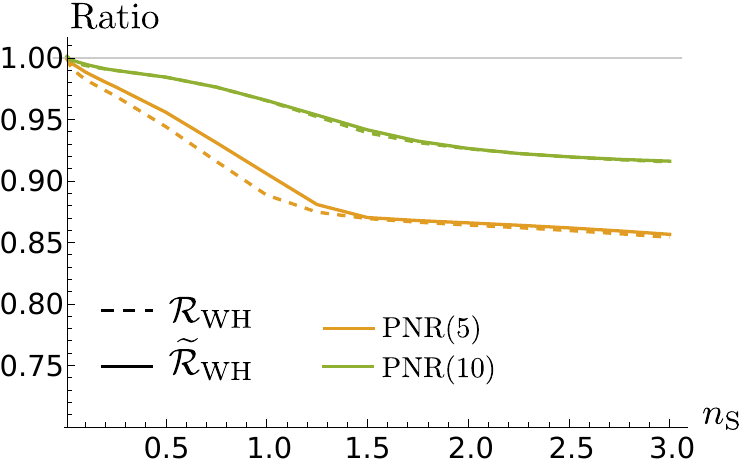} }\vspace*{2ex}
\centerline{\includegraphics[width=0.9\columnwidth]{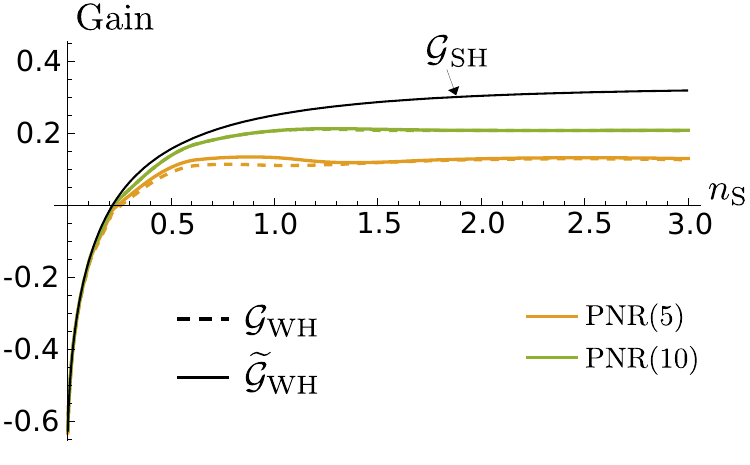} }
\caption{Plot of the gain ${\cal G}_{\WFH}$ (top panel) and ${\cal G}_{\DWFH}$ (bottom panel) as a function of the mean received energy $\ns$ for different PNR resolution $M$. ${\cal G}_{\rm SH}$ and ${\cal G}_{\rm DH}$ refers to the gain over DD achieved by single and double homodyne detection, respectively.}\label{fig:11-Opt}
\end{figure}

After transmission, Bob performs $\WFH$ detection of quadrature $q$ on the received pulses, with associated probabilities $p_{B|A}({\bf n}_B|x_A)= P^{(0)}_{\WFH}({\bf n}_B|\alpha=x_A)$ in~(\ref{eq:WFHdistr}) and
\begin{eqnarray}
\widetilde{p}_B({\bf n}_B)= \int_{\mathbb{R}} dx_A \, {\cal P}_{\nu}(x_A) \, p_{B|A}({\bf n}_B|x_A) \, .
\end{eqnarray}
In turn, the mutual information is equal to:
\begin{eqnarray}\label{eq:InfoWFH-nonG}
\widetilde{{\cal I}}_{\WFH} = \max_{z, \nu} \,\widetilde{{\cal I}}_{\WFH}(z, \nu) \, ,
\end{eqnarray}
with
\begin{IEEEeqnarray}{lCl}
\widetilde{{\cal I}}_{\WFH} (z, \nu) &=& {\sf H} \left[\widetilde{p}_B({\bf n}_B)\right] \nonumber \\[1ex]
&&- \int_{\mathbb{R}} dx_A \, {\cal P}_{\nu}(x_A) \, {\sf H} \left[p_{B|A}({\bf n}_B|x_A)\right]  \, ,
\end{IEEEeqnarray} 
and we perform optimization over both the LO amplitude $z>0$ and the free modulation parameter $\nu \ge 1/2$. The resulting information rate is $\widetilde{{\cal I}}_{\WFH}\ge {\cal I}_{\WFH}$, and outperforms the Gaussian-modulation rate in several regimes, as emerges by looking at the optimized modulation parameter $\nu_{\rm opt}$ reported in Fig.~\ref{fig:10-nuopt} for PNR($5$) detectors. The behavior is analogous for all resolutions $M$. We have $\nu_{\rm opt}=1/2$ for $\ns\ll 1$ and $\ns\gg 1$, showing GM to be optimal both in the photon starved and high energy regimes. On the contrary, for $10^{-2} \lesssim \ns\lesssim 1$ we identify two regimes. At first, we have $\nu_{\rm opt} \rightarrow \infty$, proving the discrete BPSK modulation to outperform continuous modulation schemes; thereafter $\nu_{\rm opt}$ becomes a decreasing function of $\ns$ and, ultimately, jumps back to $\nu=1/2$ for high enough energy. Therefore, in the intermediate energy regime employing NGM enhances the achievable mutual information.

The enhancement is more accentuated for low values of the resolution $M$, as we see by computing the ratio $\widetilde{\cal R}_{\WFH}=\widetilde{\cal I}_{\WFH}/C_{\rm SH}$ and the gain $\widetilde{\cal G}_{\WFH}=\widetilde{\cal I}_{\WFH}/C_{\rm DD}-1$, plotted in the top and bottom panel of Fig.~\ref{fig:11-Opt}, respectively. In particular, 
for $M=5,10$ we have an enhancement of $1.95\%$ and $0.02\%$ for the ratio, while of $19\%$ and $1.5\%$ for the gain. In fact, for larger $M$, $\WFH$ measurement approaches standard homodyne detection, thus the performance of a Gaussian input coherent ensemble, being in general suboptimal, becomes closer to the optimal one.

\section{Conclusion \label{sec: Concl}}
In this paper we investigated the role of weak-field measurements, realized by a homodyne setup employing PNR($M$) detectors and low-intensity LO, for quantum communications over a lossy bosonic channel employing a coherent state ensemble. 
In particular, we considered three relevant schemes, namely $\WFH$, $\HL$ and $\DWFH$ detection. The former two falls under the single quadrature measurement scenario and only differ by the access to the photon-number statistics on the two output beams of the homodyne scheme. 
Instead, $\DWFH$ realizes double quadrature weak-field detection via joint $\WFH$ detection of the two field quadratures.

To begin with, we considered a Gaussian modulation of coherent states at the source and computed the mutual information associated with single-quadrature measurement, comparing the $\WFH$ and $\HL$ schemes.
We proved $\WFH$ to outperform $\HL$ for all values of the mean signal energy, as a consequence of the data processing inequality. Remarkably, we showed that both $\WFH$ and $\HL$ provide an enhancement over Shannon homodyne capacity in the photon starved regime $\ns\ll 1$, whilst becoming suboptimal for $\ns \gtrsim 10^{-3}$.
The enhancement is obtained by suitably optimizing the LO intensity of the weak-field scheme exploiting the information about the mean signal energy, i.e. $z^2=z^2(\ns)$.  This information makes the weak-field probability distribution significantly different from the standard homodyne one, resulting in a higher achievable information rate when $\ns \ll z^2(\ns) \lesssim 1$.

Thereafter, we extended the former analysis to double weak-field quadrature detection, proving $\DWFH$ employing a signal-dependent optimized LO to beat both Shannon capacities in the photon starved limit.

Finally, we considered a non-Gaussian input modulation, by adopting a Gamma distribution of the coherent signal energies, and obtained an increase in the mutual information in the regime $10^{-2} \lesssim \ns \lesssim 1$, whilst showing the optimality of Gaussian modulation both in the photon starved and the high energy regimes.

The results obtained in the paper offer a glimpse into the role of weak-field measurements for quantum communications and demonstrate the choice of the LO as a key element to enhance the detection setup, raising the interesting problem of computing the channel capacity with optimized LO \cite{Lukanowski2021}.

Furthermore, our findings suggest $\WFH$ as a possible resource for protocols operating in photon starved limit, ranging from deep space communications \cite{Banaszek2019, Sidhu2021, Jarzyna2024} to optical key distribution \cite{Ikuta2016, BanaszekJachura2021, Jarzyna2023}.
More generally, they pave the way for the design of new feasible hybrid quantum receivers, employing the technologies of both discrete and continuous variable detection schemes, aiming at closing the gap between Shannon and Holevo capacities.



\ifCLASSOPTIONcaptionsoff
  \newpage
\fi




\begin{thebibliography}{10}
\baselineskip 12pt
\providecommand{\url}[1]{#1}
\csname url@samestyle\endcsname
\providecommand{\newblock}{\relax}
\providecommand{\bibinfo}[2]{#2}
\providecommand{\BIBentrySTDinterwordspacing}{\spaceskip=0pt\relax}
\providecommand{\BIBentryALTinterwordstretchfactor}{4}
\providecommand{\BIBentryALTinterwordspacing}{\spaceskip=\fontdimen2\font plus
\BIBentryALTinterwordstretchfactor\fontdimen3\font minus
  \fontdimen4\font\relax}
\providecommand{\BIBforeignlanguage}[2]{{%
\expandafter\ifx\csname l@#1\endcsname\relax
\typeout{** WARNING: IEEEtran.bst: No hyphenation pattern has been}%
\typeout{** loaded for the language `#1'. Using the pattern for}%
\typeout{** the default language instead.}%
\else
\language=\csname l@#1\endcsname
\fi
#2}}
\providecommand{\BIBdecl}{\relax}
\BIBdecl


\bibitem{Cover2006}
T.~M. Cover and J.~A. Thomas, \emph{Elements of Information Theory},
  2nd~ed.\hskip 1em plus 0.5em minus 0.4em\relax Hoboken, NJ: Wiley, 2006.

\bibitem{Cariolaro2015}
G. Cariolaro, \emph{Quantum communications}.\hskip 1em plus 0.5em minus
  0.4em\relax Springer, 2015.

\bibitem{Holevo2011}
A.~S. Holevo and V. Giovannetti, 
``Quantum channels and their entropic characteristics", 
{\em Rep. Prog. Phys.}, vol.~75,  p. 046001, 2012.

\bibitem{Banaszek2020}
K. Banaszek, L. Kunz, M. Jachura, and M. Jarzyna, 
``Quantum limits in optical communications,'' 
\emph{J. Light. Technol.}, vol.~38, no.~10, pp. 2741--2754, 2020.

\bibitem{Hemmati2011}
H.~Hemmati, A.~Biswas, and I.~B. Djordjevic,
``Deep-space optical
  communications: Future perspectives and applications,'' 
\emph{Proc. IEEE}, vol.~99, no.~11, pp. 2020--2039, 2011.

\bibitem{Gordon1962}
J.~P. {Gordon}, ``Quantum effects in communications systems,''
\emph{Proc. IRE}, vol.~50, no.~9, pp. 1898--1908, 1962.

\bibitem{Yamamoto1986}
Y.~Yamamoto and H.~A. Haus, 
``Preparation, measurement and information capacity
  of optical quantum states,'' \emph{Rev. Mod. Phys.}, vol.~58, pp. 1001--1020, 1986.

\bibitem{Caves1994}
C.~M. Caves and P.~D. Drummond, 
``Quantum limits on bosonic communication   rates,'' 
\emph{Rev. Mod. Phys.}, vol.~66, pp. 481--537, 1994.

\bibitem{Chen2012}
J.~Chen, J.~L. Habif, Z.~Dutton, R.~Lazarus, and S.~Guha, 
``Optical codeword
  demodulation with error rates below the standard quantum limit using a
  conditional nulling receiver,'' 
\emph{Nat. Photonics}, vol.~6, pp. 374--379,  2012.

\bibitem{DiMario2019}
M.~T. DiMario, L. Kunz, K. Banaszek, and F.~E. Becerra,
``Optimized communication strategies with binary coherent states over phase noise channels,"
{\em npj Quantum Inf.}, vol.~5, Art.~no.~65, 2019.

\bibitem{Notarnicola2023:KB}
M.~N. Notarnicola, M. Jarzyna, S. Olivares, and K. Banaszek,
``Optimizing state-discrimination receivers for continuous-variable quantum key distribution over a wiretap channel,"
{\em New J. Phys.}, vol.~25, Art.~no.~103014, 2023.

\bibitem{Bowen1968}
J.~I. Bowen, ``The effect of attenuation on the capacity of a photon channel,''
  \emph{IEEE Trans. Inf. Theor.}, vol.~14, no.~1, pp. 44--50, 1968.

\bibitem{Shamai1990}
S.~Shamai~(Shitz), ``On the capacity of a pulse amplitude modulated direct
  detection photon channel,'' \emph{Proc. Inst. Elec. Eng.}, vol. 137, no.~6,
  pp. 424--430, 1990.

\bibitem{Martinez2007}
A. Martinez, 
``Spectral efficiency of optical direct detection,''
 \emph{J. Opt. Soc. Am. B}, vol.~24, no.~4, pp. 739--749, 2007.

\bibitem{Cheraghchi2019}
M.~Cheraghchi and J.~Ribeiro, 
``Improved upper bounds and structural results on   the capacity of the discrete-time {P}oisson channel,'' 
\emph{IEEE Trans. Inf.  Theor.}, vol.~65, no.~7, pp. 4052--4068, 2019.

\bibitem{Lukanowski2021}
K. Łukanowski and M. Jarzyna, 
``Capacity of a Lossy Photon Channel With Direct Detection,"
{\em IEEE Trans. Commun.}, vol. 69, no. 8, pp. 5059-5068, 2021.

\bibitem{Donati2014}
G.~Donati, T.~J.~Bartley, X.-M.~Jin, M.-D.~Vidrighin, A.~Datta, M.~Barbieri and I.~A.~Walmsley, 
``Observing optical coherence across Fock layers with weak-field homodyne detectors,"
{\em Nat. Commun.}, vol.~5, Art.~no.~5584, 2014.

\bibitem{Thekkadath2020}
G.~S.~Thekkadath et al.,
``Tuning between photon-number and quadrature measurements with weak-field homodyne detection,"
{\em Phys. Rev. A}, vol.~101, Art.~no.~031801(R), 2020.

\bibitem{Allevi2017}
A. Allevi, M. Bina, S. Olivares and M. Bondani, 
``Homodyne-like detection scheme based on photon-number-resolving detectors,"
{\em Int. J. Quantum Inf.}, vol.~15, no.~8, Art.~no.~1740016, 2017.

\bibitem{Olivares2019}
S.~Olivares, A.~Allevi, G.~Caiazzo, M.~G.~A.~Paris and M.~Bondani,
``Quantum tomography of light states by photon-number-resolving detectors,"
{\em New J. Phys.} vol.~21, Art.~no.~103045, 2019.

\bibitem{Chesi2019}
G.~Chesi, L.~Malinverno, A.~Allevi, R.~Santoro, M.~Caccia, A.~Martemiyanov and M.~Bondani, 
``Optimizing Silicon photomultipliers for Quantum Optics,"
{\em Sci. Rep.} vol.~9, Art.~no.~7433, 2019.

\bibitem{Notarnicola2023:HYNORE}
M.~N.~Notarnicola, M.~G.~A.~Paris, and S.~Olivares, 
``Hybrid near-optimum binary receiver with realistic photon-number-resolving detectors," 
\emph{J. Opt. Soc. Am. B}, vol.~40, pp.~705-714, 2023. 

\bibitem{Notarnicola2023:FF}
M.~N.~Notarnicola and S.~Olivares, 
``Beating the standard quantum limit for binary phase-shift-keying discrimination with a realistic hybrid feed-forward receiver," 
\emph{Phys. Rev. A}, vol.~108, Art.~no.~042619, 2023. 

\bibitem{Notarnicola2024:PHN}
M.~N.~Notarnicola and S.~Olivares, 
``A robust hybrid receiver for binary phase-shift keying discrimination in the presence of phase noise,"
arXiv:2312.16493 [quant-ph], 2023.

\bibitem{Cattaneo2018}
M. Cattaneo, M.~G.~A. Paris, and S. Olivares,
``Hybrid quantum key distribution using coherent states and photon-number-resolving detectors,"
{\em Phys. Rev. A}, vol.~98, Art.~no.~012333, 2018.

\bibitem{Paris2012}
M.~G.~A. Paris,
``The modern tools of quantum mechanics",
{\em Eur. Phys. J. ST} vol.~203, Art.~no.~61, 2012.

\bibitem{Holevo1973}
A.~S. Holevo, 
``Bounds for the quantity of information transmitted by a quantum
  communication channel,'' 
\emph{Probl. Inf. Transm.}, vol.~9,   no.~3, pp. 177--183, 1973.

\bibitem{Holevo1998}
A.~S. Holevo, 
``The capacity of the quantum channel with general signal states,'' 
\emph{IEEE Trans. Inf. Theory}, vol.~44, no.~1, pp.~269--273, 1998.

\bibitem{Giovannetti2014}
V.~Giovannetti, R.~Garc{\'\i}a-Patr{\'o}n, N.~J. Cerf, and A.~S. Holevo,
  ``Ultimate classical communication rates of quantum optical channels,''
  \emph{Nature Photon.}, vol.~8, pp. 796--800, 2014.

\bibitem{Olivares2021}
S.~Olivares, ``Introduction to generation, manipulation and characterization of
  optical quantum states,'' \emph{Phys. Lett. A}, vol.~418, Art.~no.~127720, 2021.

\bibitem{Holevo2001}
A.~S. Holevo and R.~F. Werner, ``Evaluating capacities of bosonic {G}aussian
  channels,'' \emph{Phys. Rev. A}, vol.~63, p. 032312, 2001.

\bibitem{Giovannetti2004}
V.~Giovannetti, S.~Guha, S.~Lloyd, L.~Maccone, J.~H. Shapiro, and H.~P. Yuen,
  ``Classical capacity of the lossy bosonic channel: The exact solution,''
  \emph{Phys. Rev. Lett.}, vol.~92, p. 027902, 2004.

\bibitem{Shannon1949}
C.~E. Shannon, ``Communication in the presence of noise,'' \emph{Proc. IRE},
  vol.~37, no.~1, pp. 10--21, 1949.

\bibitem{Guha2011}
S.~Guha, 
``Structured optical receivers to attain superadditive capacity and the Holevo limit,'' 
\emph{Phys. Rev. Lett.}, vol. 106, no.~24, p. 240502, 2011.

\bibitem{Banaszek2017}
K.~Banaszek and M.~Jachura, 
``Structured optical receivers for efficient   deep-space communication,'' 
in \emph{2017 IEEE International Conference on   Space Optical Systems and Applications (ICSOS)}, 2017, pp. 34--37.

\bibitem{Olivares2020}
S. Olivares, A. Allevi and M. Bondani
``On the role of the local oscillator intensity in optical homodyne-like tomography,'' 
{\em Phys. Lett. A}, vol.~384, no.~17, Art.~no.~126354, 2020.

\bibitem{Blahut1972}
R.~Blahut, ``Computation of channel capacity and rate-distortion functions,''
  \emph{IEEE Trans. Inf. Theor.}, vol.~18, no.~4, p. 460–473, 1972.

\bibitem{Arimoto1972}
S.~Arimoto, ``An algorithm for computing the capacity of arbitrary discrete
  memoryless channels,'' \emph{IEEE Trans. Inf. Theor.}, vol.~18, no.~1, pp.
  14--20, 1972.

\bibitem{Dolinar2011}
S.Dolinar, K.~M. Birnbaum, B.~I. Erkmen, and B. Moision, 
``On approaching the ultimate limits of photon-efficient and bandwidth-efficient optical communication," 
in {\em Proc. IEEE Int. Conf. Space Opt. Syst. Appl.}, pp.~269--278, 2011.

\bibitem{Hogg2019}
R.~V Hogg, J.~W. McKean, and A.~T. Craig,
{\em Introduction to Mathematical Statistics}. Pearson, 2019.

\bibitem{Banaszek2019}
K. Banaszek, L. Kunz, M. Jarzyna, and M. Jachura,
``Approaching the ultimate capacity limit in deep-space optical communication," 
in {\em Proc. SPIE} 10910, Free-Space Laser Communications XXXI, 109100A, 2019.

\bibitem{Sidhu2021}
J.~S. Sidhu, et al.,
``Advances in space quantum communications,"
{\em IET Quant. Comm.}, vol.~2, no.~4, pp.~182–217, 2021.

\bibitem{Jarzyna2024}
M. Jarzyna, L. Kunz, W.Zwoliński, M. Jachura, and K. Banaszek,
``Photon information efficiency limits in deep-space optical communications,,"
{\em Opt. Eng.}, vol.~63, Art.~no.~041209, 2024. 

\bibitem{Ikuta2016}
T. Ikuta and K. Inoue, 
``Intensity modulation and direct detection quantum key distribution based on quantum noise,” 
{\em New J. Phys.}, vol.~18, no.~1, Art.~no.~013018, 2016.

\bibitem{BanaszekJachura2021}
K.~Banaszek, M.~Jachura, P~Kolenderski and M.~Lasota, 
``Optimization of intensity-modulation/direct-detection optical key distribution under passive eavesdropping,"
{\em Opt. Express}, vol.~29, pp.~43091--43103, 2021.

\bibitem{Jarzyna2023}
M. Jarzyna, M. Jachura and K. Banaszek, 
``Quantum Pulse Gate Attack on IM/DD Optical Key Distribution Exploiting Symbol Shape Distortion," 
{\em IEEE Commun. Lett.}, vol.~27, no.~7, pp.~1699-1703, 2023.





\end{thebibliography}


%




\end{document}